\documentclass[%
 twocolumn,
 amsmath,amssymb,prc,aps,reprint,showpacs,floatfix]{revtex4-1}

\usepackage{docs}%
\usepackage{bm}%
\usepackage{graphicx,array,multirow}
\usepackage{rotating}
\usepackage{hyperref}
\usepackage{url}
\usepackage{srcltx}
\expandafter \ifx \csname package@font\endcsname\relax\else
\expandafter \expandafter \expandafter
\usepackage
\expandafter \expandafter
\expandafter{\csname package@font\endcsname}%
\fi
\usepackage{soul}

\usepackage{epstopdf}
\newcommand{\foot}[2]{\mbox{\tiny $\stackrel{#1}{#2}$}}
\newcommand{\el}{\ \\\nonumber}

\newcommand{\sh}[1]{#1\hskip-7pt \diagup}

\newcommand{\ket}[1]{\left|#1\right>}

\newcommand{\bracket}[3]{\left<#1\left|#2\right|#3\right>}

\newcommand{\dc}[1]{\delta^{(4)}( #1)}

\newcommand{\mb}[1]{\bm{#1}}

\newcommand{\ups}[2]{u_{#2}(\bm{#1})}

\newcommand{\oups}[2]{\overline{u}_{#2}(\bm{#1})}

\newcommand{\dpt}{(2\pi)^3}

\newcommand{\tr}{{\mathrm{Tr}}}

\newcommand{\gfiv}{\gamma^5}

\begin{document}
\title{Electroweak Form Factors of the $\Delta$(1232) Resonance}

\author{Krzysztof M. Graczyk, Jakub \.{Z}muda, Jan T. Sobczyk}

\affiliation{Institute of Theoretical Physics, University of Wroc\l aw, pl. M. Borna 9,
50-204, Wroc\l aw, Poland}

\date{\today}%

\begin{abstract}
Nucleon $\to \Delta(1232)$ transition electroweak form factors are discussed in a  single pion production model
with nonresonant background terms originating from a chiral perturbation theory.
Fits to electron-proton scattering $F_2$  as well as neutrino scattering bubble chamber experimental
data are performed. Both $\nu$-proton and $\nu$-neutron channel data are discussed in a unified statistical model.
A new parametrization of the $N\to\Delta(1232)$ vector form factors is proposed.
Fit to neutrino scattering data gives axial mass $M_{A\Delta}=0.85\foot{+0.09}{-0.08}\;$(GeV) and 
$C_5^A(0) = 1.10\foot{+0.15}{-0.14}$ in accordance with Goldberger-Treiman relation as long as
deuteron nuclear effects are considered.
\end{abstract}

\maketitle

\section{Introduction}
\label{sec:spp}

Weak single pion production (SPP) processes have been studied for many decades,
but their importance for the neutrino physics has grown with the development
of accelerator neutrino experiments. In the few-GeV energy range characteristic
for the experiments such as T2K \cite{Abe:2011ks}, MINOS \cite{Evans:2013pka}, NOvA \cite{Ayres:2004js}, 
MiniBooNE \cite{AguilarArevalo:2007it}, and LBNE \cite{Adams:2013qkq} this
interaction channel contributes a large fraction of the total cross section.
Rough estimates show, that for an isoscalar target and neutrino energy of
around 1 GeV SPP accounts for about 1/3 of the interactions. 

The SPP events with pion absorption contribute to the background in measurements of  quasi-elastic neutrino  scattering on nuclear targets. 
Neutral current  $\pi^0$ production processes  add to the background for the $\nu_e$ appearance measurement in water Cherenkov detectors. 
The detailed estimate of  the cross-sections for the SPP is important for a correct extraction of neutrino oscillation parameters 
in long baseline experiments.

Theoretical modelling of the SPP processes on nuclear targets suffers from extra complications.  Any
attempt to obtain an information about the  Nucleon (N) to $\Delta(1232)$ resonance transition vertex
from these data is biased by systematic errors coming from nuclear model
uncertainties. On the experimental side, there seems to be a tension between the
MiniBooNE and very recent MINER$\nu$A SPP data on
(mostly) carbon target, see Ref. \cite{Eberly:2014mra}.
For hereby analysis measurements of the neutrino-production on free or almost free targets are desired.
At present such data exist only for $\sim$30 years old Argonne National Laboratory (ANL) \cite{Barish:1978pj,Radecky:1981fn} and Brookhaven National Laboratory (BNL) 
\cite{Kitagaki:1986ct,Kitagaki:1990vs} bubble chamber experiments, where deuteron and hydrogen targets were utilized.
In this case one may hope to reduce the many-body bias in a reasonable manner with a simple theoretical ansatz \cite{AlvarezRuso:1998hi}.

In order to understand the neutrino SPP data
it is necessary to have a model of nonresonant background,
see Ref. \cite{Fogli:1979cz}. In more recent studies of weak SPP 
typically only the neutrino-proton channel $\nu_\mu + p\to  \mu^-+\pi^++p$ is discussed in detail
\cite{Sato:2003rq,Hernandez:2007qq,Barbero:2008zza,Hernandez:2010bx,Lalakulich:2010ss,Serot:2012rd}. This is a big drawback, because simple total
cross section ratio
analysis shows, that the background contribution is much larger in neutrino-neutron channels. The neutrino-proton SPP 
channel can be described well within a model that contains the $\Delta(1232)$ resonance
contribution only, see e.g. Ref. \cite{Graczyk:2009qm}.  In the latter paper it was 
argued that  the $d\sigma/dQ^2$ results in \cite{Radecky:1981fn} do not include the flux normalization error. 
Incorporation into the analysis this error and also deuteron effects in both ANL and BNL experiments 
allowed for a consistent fit  for both
data sets with $C_ 5^A(0)=1.19\pm 0.08$ and $M_A=0.94 \pm 0.03$ GeV. The attempt to extract the leading $C_5^A(Q^2)$
$N\to\Delta$ form factor parameters in a model containing nonresonant background has been done in Refs. \cite{Hernandez:2007qq,Hernandez:2010bx}.
The results both for a model without \cite{Hernandez:2007qq} and with deuteron effects \cite{Hernandez:2010bx}
gave the values of $C_5^A(0)$ far from the Goldberger-Treiman relation estimate of $C_5^A(0)\approx 1.2$ \cite{Goldberger:1958tr}
($C_5^A(0)=0.867\pm 0.075$ in \cite{Hernandez:2007qq} and $C_5^A(0)=1.00\pm 0.11$ in \cite{Hernandez:2010bx}). 
From the above mentioned models only those in Refs. \cite{Sato:2003rq,Barbero:2008zza} have been directly 
validated on the electroproduction processes. Some authors use vector form factor parametrization
from Ref. \cite{Lalakulich:2006sw}, based on the MAID analysis \cite{Drechsel:2007if}. The authors of Ref. \cite{Lalakulich:2006sw} proposed a model
containing only $\Delta$ resonance contribution without any background and compared it to the MAID2007 helicity amplitudes.
The problem is that the $\Delta$  helicity amplitudes extraction procedure
is model-dependent. There are important $\Delta$ -- background interference effects and separation procedure
depends on the background model details. It is important to have $\Delta$ form factor consistent with the other
ingredients of the model.
\begin{figure*}[!htb]
\centering\includegraphics[width=0.8\textwidth]{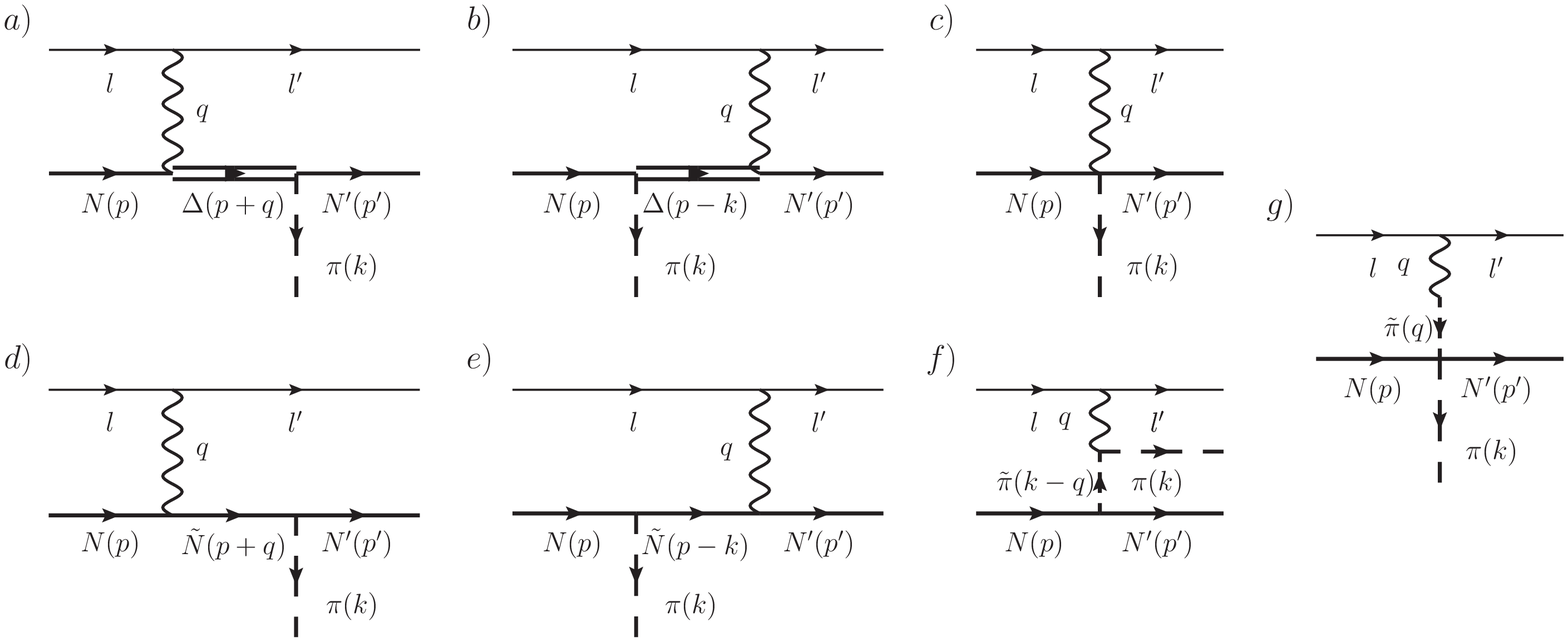}
\caption{(Color onlne) Basic pion production diagrams from \cite{Hernandez:2007qq}:
a) Delta pole ($\Delta$P), b) crossed  Delta pole (C$\Delta$P), c) contact term (CT), d) nucleon pole (NP),
e) crossed nucleon pole (CNP), f) pion-in-flight (PIF), g) pion pole (PP) }\label{fig:mec}
\end{figure*}

Keeping in mind the above caveats of previous analyses we propose an improved approach. We adapt and develop the statistical framework of
Ref. \cite{Graczyk:2009qm} in order to
fit both vector and axial form factors of the $\Delta(1232)$ resonance. We use inclusive electron-proton scattering data for the electromagnetic
interaction in the $\Delta(1232)$ region and deuteron bubble chamber data
for the weak one. For the latter we expand the previously used statistical approach in order to incorporate the neutron channels,
which was never done before.
In this manner we include the data sets, that are very sensitive to the nonresonant background.

This paper is organized as follows: section \ref{sec:genform} is devoted to the general formalism of single pion production, section
\ref{sec:statistical}
introduces the statistical model of our analysis, section \ref{sec:results} shows our main results and finally section
\ref{sec:conclusions} contains
the conclusions.

\section{General formalism}
\label{sec:genform}

We discuss the charged current inelastic neutrino scattering off nucleon targets. Three channels for neutrino SPP interactions are:
\begin{eqnarray}
\label{channel_1}
\nu_\mu(l) + p(p) & \to & \mu^-(l') + \pi^+(k) + p(p')
\\
\label{channel_2}
\nu_\mu(l) + n(p) & \to & \mu^-(l') + \pi^0(k) + p(p')
\\
\label{channel_3}
\nu_\mu(l) + n(p) & \to & \mu^-(l') + \pi^+(k) + n(p')
\end{eqnarray}
with $l$, $l'$, $p$, $p'$ and $k$ being the neutrino, muon, initial nucleon, final nucleon and pion four momenta respectively. 
The four momentum transfer is defined as:
\begin{eqnarray}
\label{eq:4momtransf}
q=l-l'=p'+k-p,\quad Q^2=-q^2,\quad q^\mu = (q^0,\mathbf{q})
\end{eqnarray}
and the square of hadronic invariant mass is:
\begin{eqnarray}
\label{eq:W}
W^2=(p+q)^2=(p'+k)^2.
\end{eqnarray}
Throughout this paper
the metric $g^{\mu\nu}=\mathrm{diag(+,-,-,-)}$ is used.

For the pion electroproduction we are interested in proton target reactions;
\begin{eqnarray}
\label{channel_1e}
e^-(l) + p(p) & \to & e^-(l') + \pi^+(k) + n(p')
\\
\label{channel_2e}
e^-(l) + p(p) & \to & e^-(l') + \pi^0(k) + p(p').
\end{eqnarray}
In the $1$~GeV energy region the process (\ref{channel_1}) is overwhelmingly dominated by the
intermediate $\Delta^{++}(1232)$ state.
The dominance of the resonant pion production mechanism makes this channel
attractive for the analysis of the $\Delta(1232)$ properties. The other two channels (Eqs. (\ref{channel_2}) and (\ref{channel_3})) are known
to have a large nonresonant pion production contribution and thus
present more challenges for theorists. 
\subsection{Cross section}

The inclusive double differential SPP cross section for neutrino scattering off nucleons at rest  has the following form:
\begin{eqnarray}\nonumber\label{eq:3diffnucleon}
\frac{d^2\sigma}{dQ^2dW}&=&\frac{4\pi^2}{E^2} G^2_F\cos^2\theta_C \frac{W}{M} \hspace{-5pt}\int \hspace{-5pt}\frac{d^3k}{\dpt 2E_\pi (k)}L_{\mu\nu}H^{\mu\nu}\\
L^{\mu\nu}&\hspace{-4pt}=\hspace{-4pt}&l^\mu l'^{\nu}+l^\nu l'^{\mu}-g^{\mu\nu}l\cdot l'+i\epsilon^{\mu\nu\alpha\beta}l'_{\alpha}l_{\beta}\el
H^{\mu\nu}&\hspace{-4pt}=\hspace{-4pt}&\frac{1}{2}\hspace{-3pt}\int \hspace{-5pt}\frac{d^3p'}{\dpt}\frac{1}{4ME (p')}\overline{\sum_{spins}}\bracket{\pi N'}{j^\mu_{cc}(0)}{N}\el&&\bracket{\pi N'}{j^\nu_{cc}(0)}{N}^\ast\dc{p'\hspace{-3pt}+ \hspace{-2pt} k \hspace{-2pt}- \hspace{-2pt}p \hspace{-2pt}- \hspace{-2pt}q}=\el &\hspace{-4pt}=\hspace{-4pt}&\frac{1}{128\pi^3 M E(p')} \hspace{-2pt}A^{\mu\nu}\delta ( E(p')\hspace{-2pt}+ \hspace{-2pt}E_\pi (k) \hspace{-2pt} -\hspace{-2pt} M\hspace{-2pt} -\hspace{-2pt} q^0),
\end{eqnarray}
where $E$ is incident neutrino energy, $M$ is the averaged nucleon mass, $E_\pi(k)$ and $E(p')$ are the final state pion and nucleon energies, 
$G_F=1.1664\cdot 10^{-11}\;$MeV$\mathrm{^{-2}}$ is the Fermi constant, $L^{\mu\nu}$ - the leptonic and $H^{\mu\nu}$ - the hadronic tensors. 
The Cabibbo angle, $\cos(\theta_C)=0.974$,  was factored out of the weak charged current definition.

The information about dynamics of SPP is contained in matrix elements, $\bracket{\pi N'}{j^\mu_{cc}(0)}{N}$, 
which describe the transition between an
initial nucleon state $\ket{N}$ and a final nucleon-pion state $\ket{\pi N'}$. One can introduce ``reduced current matrix elements''
$s^\mu$ and express the weak transition amplitudes:
\begin{eqnarray}
\label{eq:smu}
\bracket{N'(p',s')\pi(k)}{j^\mu_{cc}}{N(p,s)}=\oups{p'}{s'}s^\mu\ups{p}{s},
\end{eqnarray}
with isospin information hidden inside $s^\mu$.

After performing the summations over nucleon spins we can rewrite the hadronic tensor as:
\begin{eqnarray}
\label{eq:amunu1p1h1pi}
A^{\mu\nu}&=&\tr\left[(\sh{p}'+M)s^\mu (\sh{p}+M)\gamma^0s^{\nu\dag}\gamma^0\right]
\end{eqnarray}
where $\sh{p} = \gamma_\mu p^\mu $.  

The differential cross section on free nucleons becomes then:
\begin{eqnarray}\label{eq:dq2free}
\frac{d^2\sigma}{dWdQ^2}&=&\frac{G_F^2\cos^2(\Theta_C)}{512\pi^4 E^2 M}
\int\hspace{-3pt} d\Omega_\pi \hspace{-3pt}\int_0^{\infty} \hspace{-8pt} \frac{\mathbf{k}^2 d|k|}{E_\pi(k)
E(p')}  \el& & \frac{W}{M} L_{\mu\nu}A^{\mu\nu} \delta ( E(p')\hspace{-2pt}+\hspace{-2pt}E_\pi (k)\hspace{-2pt} -\hspace{-2pt} M\hspace{-2pt} -\hspace{-2pt} q^0).
\end{eqnarray}

In the model of this paper the dynamics of SPP process is defined by a set of Feynman diagrams (Fig. \ref{fig:mec})
with vertices determined by the effective chiral field theory. They are discussed in Ref. \cite{Hernandez:2007qq}, 
where one can find exact expressions for $s^\mu$.
The same set of diagrams describes also pion electroproduction, with the exception of the pion pole diagram,
which is purely axial. We call this approach "HNV model" after the names of the authors of Ref. \cite{Hernandez:2007qq}.

\subsection{$N\to\Delta(1232)$ excitation}
\label{sec:delta}

The $\Delta(1232)$ resonance excitation is treated within the isobar framework.
For positive parity spin-$\frac{3}{2}$ particles we can write down a general form of the electroweak excitation vertex:
\begin{equation}
\nonumber
\label{eq:delver}
\Gamma^{\alpha\mu}(p,q)
= \left[V^{\alpha\mu}_{3/2}- A^{\alpha\mu}_{3/2}\right]\gfiv
\end{equation}
where
\begin{widetext}
 \begin{eqnarray}
 V^{\alpha\mu}_{3/2} &=&  \frac{C^V_3(Q^2)}{M}(g^{\alpha\mu}\sh{q}  -
q^\alpha\gamma^\mu )   +   \frac{C^V_4(Q^2)}{M^2}(g^{\alpha\mu}q  \cdot  (p  +  q)  -  q^\alpha (p  +  q)^\mu)  + 
\frac{C^V_5(Q^2)}{M^2}(g^{\alpha\mu}q  \cdot  p -  q^\alpha p^\mu )  +  g^{\alpha\mu} C_6^V(Q^2) \nonumber \\
\\
-A^{\alpha\mu}_{3/2} & = & \left[ \frac{C^A_3(Q^2)}{M}(g^{\alpha\mu}\sh{q}  -  q^\alpha\gamma^\mu )  +  \frac{C^A_4(Q^2)}{M^2}(g^{\alpha\mu}q  
\cdot  (p  +  q)  -
q^\alpha (p  +  q)^\mu) +  C_5^A(Q^2) g^{\alpha\mu}  +  \frac{C^A_6(Q^2)}{M^2}q^\alpha q^\mu\right] \gfiv.
\end{eqnarray}
\end{widetext}

A relevant information about the inner structure of the $\Delta(1232)$ resonance is contained in a set of vector and axial form factors,
$C_j^{V,A}$ assumed to be functions
of $Q^2$ only (with the exception of $C_4^V$ which depends also on $W$).

\subsection{Conserved vector current and vector form factors}

Thanks to conserved vector current (CVC) hypothesis we can express weak vector form factors by electromagnetic ones.
There exist several parametrizations of $C_j^V$ proposed over the course of past five decades,
see Refs. \cite{Jones:1972ky,Dufner:1967yj,Lalakulich:2006sw,Drechsel:2007if}.
In this paper we propose our own model in order to be consistent with the chosen description of the nonresonant background.
The size and excellent accuracy of the electromagnetic data set allows for an introduction of multiple fit parameters.
 
We assume that the $N\to \Delta$ transition form factors have the same large $Q^2$  behaviour
as the electromagnetic elastic nucleon form factors. The theoretical arguments 
 \cite{Brodsky:1974vy} suggest that at $Q^2 \to \infty $ the nucleon form factors fall down as $1/Q^4$ and we adopt
appropriate Pad\'{e} type parametrization \cite{Kelly:2004hm}.
We allow for a violation from the $SU(6)$-symmetry quark model relations $C_4^V(Q^2)=-\frac{M}{W}C_3^V(Q^2)$ 
and $C_5^V=0$ between the form factors \cite{Liu:1995bu}.  Finally, to reduce the number of parameters in $C_5^V$ 
we assume  the  dipole representation. Altogether, our parametrization has the following form:
\begin{eqnarray}
\label{eq:Kellyc5}
C_3^V(Q^2)&=&\frac{C_3^V(0)}{1+AQ^2+BQ^4+CQ^6}\cdot(1+K_1Q^2)\\
C_4^V(Q^2)&=&-\frac{M_p}{W}C_3^V(Q^2)\cdot\frac{1+K_2Q^2}{1+K_1Q^2}\\
\label{eq:c5K} C_5^V(Q^2)&=&\frac{C_5^V(0)}{\left(1+D\frac{Q^2}{M_V^2}\right)^2}.
\end{eqnarray}
We use the standard value of the vector mass $M_V=0.84\;$GeV.
This parametrization reproduces quark model relation between $C_3^V$ and $C_4^V$ at $Q^2=0$ and is consistent with nonzero $S_{1/2}$ helicity
amplitude.

In Sect. \ref{sec:results} we present the best fit values of parameters $C_3^V(0)$, $C_5^V(0)$, $A$, $B$, $C$, $D$, $K_1$ and $K_2$.

\subsection{Partially conserved axial current and axial form factors}

In the axial part the leading contribution comes from $C_5^A(Q^2)$ which is an analogue of the isovector nucleon axial form factor.
Partially conserved axial current (PCAC) hypothesis relates the value of $C_5^A(0)$
with the strong  coupling constant $f^\ast$ through off-diagonal Goldberger-Treiman relation \cite{Goldberger:1958tr}:
\begin{eqnarray}
\label{eq:GolTreq}
C_5^A(0)=\frac{f^\ast}{\sqrt{2}}\approx 1.2,
\end{eqnarray}
but we will treat $C_5^A(0)$ as a free parameter.
Most often it is assumed, that $C_5^A$ has a dipole $Q^2$ dependence:
\begin{eqnarray}
\label{eq:dipolec5a}
C_5^A(Q^2)=\frac{C^A_5(0)}{(1+Q^2/M_{ A\Delta}^2)^2}
\end{eqnarray}
The axial mass parameter $M_{A\Delta}$ is expected to be of the order of $1\, \mathrm{ GeV}$. 
The authors of Refs. \cite{Hernandez:2007qq} and \cite{Lalakulich:2010ss}
use the parametrization of $C_5^A(Q^2)$ proposed in Ref. \cite{Paschos:2003qr}:
\begin{eqnarray}
C_5^A(Q^2)&=&\frac{C_5^A(0)}{(1+Q^2/M_{A\Delta}^2)^2}\frac{1}{(1+Q^2/(3M_{A\Delta}^2))^2}.
\end{eqnarray}
Other groups, e.g. authors of Ref. \cite{SajjadAthar:2009rd}, occasionally use parametrization from Ref. \cite{Schreiner:1973mj},
which contains even more free parameters. In our fits we assume the dipole form of $C_5^A$.

The $C_6^A$ form factor is an analogue of the nucleon induced pseudoscalar form factor. It can be related to $C_5^A$ as:
\begin{eqnarray}
\label{eq:c6a}
C_6^A(Q^2)=\frac{M^2}{m_\pi^2+Q^2}C_5^A(Q^2),
\end{eqnarray}
where $m_\pi$ is average pion mass.
The $C_3^A(Q^2)$ is the axial counterpart of the very small electric quadrupole (E2) transition form factor and
we set $C_3^A=0$. For the $C_4^A$ we use the Adler model relation \cite{Adler}:
\begin{eqnarray}
\label{eq:c4a}
C_4^A(Q^2)=-\frac{1}{4}C_5^A(Q^2).
\end{eqnarray}
In this way the axial contribution is fully determined by $C_5^A(Q^2)$.
Altogether there are two free parameters: $C_5^A(0)$ and $M_{A\Delta}$.
If there were enough experimental data one could drop the Adler relation and treat $C_4^A(Q^2)$ as an independent form factor. 
However, the  ANL and BNL experimental data do not have sufficient statistics to obtain separate fits of $C_5^A$ and $C_4^A$ \cite{Graczyk:2009zh}, 
see also the discussion in Ref. \cite{Hernandez:2010bx}.  

\subsection{Deuteron effects}

In this paper we
consider a deuteron model based on phenomenological nucleon momentum distribution. The following effects are taken into account:

\begin{itemize}
 \item Nucleon momentum distribution $f(p)$ taken from the Paris potential \cite{Lacombe:1981eg} (also used by the authors of 
 Ref. \cite{Hernandez:2010bx}). We verified that
 other parameterizations (Hulthen \cite{Hulthen}, Bonn \cite{Machleidt:1987hj}) lead to very similar results.
 \item {\it Flux correction} coming from varying relative neutrino-nucleon velocity:
 \begin{equation}
\label{eq:vrel}
v_{rel.} =  \frac{\sqrt{(l\cdot p)^2}}{E E(p)} = \left| \frac{(E E(p) \hspace{-1pt}-\hspace{-1pt} \mathbf{l}\cdot \mathbf{p})}{EE(p)} \right|
= \left| 1 \hspace{-1pt}- \hspace{-1pt}\frac{\mathbf{l}\cdot \mathbf{p}}{EE(p)} \right|.
\end{equation}
\item Realistic energy balance within plane wave impulse approximation (PWIA). 
It is assumed that the spectator nucleon does not participate in the interaction.
In the case of quasielastic neutrino scattering it was shown in \cite{Shen:2012xz} that 
for neutrino energies larger than $500$~MeV final state interactions effects violating PWIA are very small. The effective, momentum
dependent, binding energy becomes:
\begin{equation}
B(p) =2 E(p) - M_D,
\end{equation}
where $M_D$ is deuteron mass.
\item De Forest treatment of the off-shell matrix elements \cite{DeForest:1983vc}.
\end{itemize}
The expression for the cross section becomes:
\begin{widetext}
\begin{eqnarray}
\label{eq:dq2deut}
\frac{d\sigma}{dQ^2 dW}&=& \int d^3p
\frac{f(p)}{v_{rel.}} \frac{G_F^2\cos^2(\Theta_C)|\mb{l'}|}{16\pi E_\nu E(p)|\mathcal{J}|}\int\frac{d^3k}
{\dpt 2E_\pi(k)}\int\frac{d^3p'}{\dpt 2E(p')} L_{\mu\nu}A^{\mu\nu}(p,\tilde{q},k)\delta^4(p+\tilde{q}-k-p').
\end{eqnarray}
\end{widetext}
with
$\tilde q^\mu = (q^0-B(p), \vec q)$ and 
\begin{eqnarray}\label{eq:jac}
\mathcal{J}&=&Det\left(\begin{array}{cc} \frac{\partial Q^2}{\partial \cos(\Theta)} & \frac{\partial Q^2}{\partial q^0}\\ 
\frac{\partial W}{\partial \cos(\Theta)} & \frac{\partial W}{\partial q^0} \end{array}\right).
\end{eqnarray}
The explicit form of the Jacobian $\mathcal{J}$ is complicated because the 
invariant mass $W$ depends both on the energy transfer $q^0$ and the lepton scattering angle $\Theta$.
\section{Statistical framework}
\label{sec:statistical}
Our main goal is to have a SPP model working for the weak pion production.
A natural procedure is to 
extract the information about vector and axial form factors independently using first respective electron scattering 
and then neutrino SPP data. In the next
paragraphs we describe details of our statistical model.

\subsection{Vector Contribution to Weak SPP}
\label{sec:fitosip}

The available electron data set is very prolific and accurate compared to the neutrino data. 
One can extract the information about the functional form of the vector $N\to \Delta$ transition
form factors from several observables, including electron/target polarizations. 
Dedicated electroproduction experiments were performed in JLab and Bonn
\cite{Ungaro:2006df,Joo:2001tw,Joo:2003uc,Joo:2004mi,Egiyan:2006ks}.
Our main goal is (due to a poor quality of the neutrino SPP data) to reproduce correctly only
the most important characteristics of the neutrino SPP reactions: overall cross sections and distributions in $Q^2$.
Detailed analysis of the electroproduction data should focus on pion angular distributions but it goes
beyond the scope of this paper and is going to be a subject of further studies.

We explore the information contained in  electron-proton $F_2$ data from \cite{Osipenko:2003ua}.  
In our fit we include 37 separate series (for different $Q^2$ values)
of $F_2$ data points. Since our final analysis aims at neutrino ANL experiment we have restricted ourselves to data points
from the lowest value of $Q^2$ (0.225 $\mathrm{GeV^2}$) up to
2.025 $\mathrm{GeV^2}$ only.

The data are for the inclusive structure function, thus we have limited ourselves to values of invariant mass $W$ up to $M_p+2m_\pi$. 
Beyond that value the experimental data
include more inelastic channels, starting from two pion production. Even with this limitation for $Q^2\leq 2.025$ and $W<M_p+2m_\pi$
there are still $603$ data points.

In order to ensure that the results will reproduce well the data at the $\Delta(1232)$ peak we decided to expanded our fit to $W=1.27$~GeV.
Because there are no exclusive electron SPP data in the region $W\in(M+2m_\pi ,1.27$~GeV) we 
add to our fit a term in which MAID 2007 model predictions are taken as $228$ fake data points.
The total errors are identical with those of respective Osipenko \textit{et al.} \cite{Osipenko:2003ua} points.
Additional points help to reproduce better the $\Delta(1232)$ peak region. From technical reasons
we could not apply the MAID model directly in our fits (the exact formulas
for their SPP amplitudes have never been published). We have generated these additional
points using the on-line version of MAID (\url{http://wwwkph.kph.uni-mainz.de/MAID//}).
We have also used an information about MAID 2007 model helicity amplitudes.
The caveat is that the experimental results
contain both resonant and nonresonant contributions (see e.g. Ref. \cite{Davidson:1991xz} ).
Thus the measured helicity amplitudes depend on how one defines
the "Delta" and "background".
The HNV model differs with MAID in the treatment of both and one cannot expect the extracted helicity amplitudes to be the same. The information
about helicity amplitudes enters our estimator with a large \textit{ad hoc} error assumption.

\subsection{Axial Contribution to weak SPP: neutrino bubble chamber experiments}
\label{sec:deuter}

We consider a statistical framework, proposed in Ref. \cite{Graczyk:2009qm}, which incorporates the relevant data from the ANL experiment
and allows for a treatment of both $C_5^A(0)$ and $M_{A\Delta}$ as free parameters.

The main results of the ANL experiment were published in Refs. \cite{Barish:1978pj,Radecky:1981fn}.
ANL used a neutrino beam with mean energy below 1 GeV and a large flux normalization uncertainty $\Delta p_{ANL}\sim 20\%$
that was not included in the published $d\sigma/dQ^2$ cross section for the reaction in Eq. (\ref{channel_1}) \cite{Graczyk:2009qm}.
ANL reported the data with the invariant mass cut $W<1.4$ GeV, which allows
to confine to the $\Delta(1232)$ region and neglect contributions from heavier resonances, whose axial couplings are by large unknown.
Our analysis uses information from both proton and neutron SPP channels.

In the $\nu_\mu + p \to \mu^- + p + \pi^+$  channel (denoted as A1) there are data on flux averaged differential cross section $\sigma_i^{exp}$
with respect to $Q^2$. The ANL papers provide their errors, $\Delta\sigma_i^{exp}$. By looking at the corresponding
numbers of detected events one can show that $\Delta\sigma_i^{exp}$ are statistical errors only.
Following \cite{Graczyk:2009qm} we explore this fact and make the analysis more complete by considering also  
a correlated error coming from the overall
flux normalization uncertainty. We define the $\chi^2$ estimator as:
\begin{equation}
\label{eq:chi2anl1}
\chi^2_{A1}=\sum_{i=1}^{9} \left(\frac{\sigma_i^{th}-p_{ANL}
\cdot\sigma_i^{exp}}{p_{ANL}\,\Delta\sigma_i^{exp}}\right)^2+ \left(\frac{p_{ANL} - 1}{\Delta p_{ANL}}\right)^2
\end{equation}
with ANL normalization factor $p_{ANL}$ treated as a free parameter.

The theoretical cross sections are defined as:
\begin{widetext}
\begin{eqnarray}\label{eq:sigmath}
\sigma_i^{th}&=&\frac{1}{\Delta Q^2_i} \frac{1}{ \displaystyle \int_{E_{min}}^{E_{max}} \Phi(E')dE'}
\int_{Q^2_i-\Delta Q^2_i/2}^{Q^2_i+\Delta Q^2_i/2}dQ^2\int_{E_{min}}^{E_{max}} dE
\displaystyle \Phi (E)\cdot \;  \frac{\displaystyle d\sigma^{th}(E,Q^2)}{dQ^2},
\end{eqnarray}
\end{widetext}

\begin{eqnarray}
 \frac{d\sigma^{th}(E,Q^2)}{dQ^2}=\int_{M+m_\pi}^{1.4~\mathrm{GeV}}dW\frac{d^2\sigma^{th}(E,Q^2)}{dWdQ^2}.
\end{eqnarray}
$Q^2_i$ is the $i$-th bin central $Q^2$ value, $\Delta Q_i^2$- the bin width and
$\Phi(E)$ is the ANL flux.
In this channel the integral spans neutrino energies between $E_{min}=0.5\;$GeV and $E_{max}=6\;$GeV.

For both ANL neutron channels $\nu_\mu + n \to \mu^- + p  +\pi^0$ 
(denoted as A2) and $\nu_\mu + n \to \mu^- + n + \pi^+$ (denoted as A3) the data are in a form of 
event distributions in $Q^2$ denoted as $N_i^{EXP}$ and also a few overall cross sections points. 
In our study we include experimental correction factors $C^{exp}$, $N_j^{exp} \to C^{exp} N_j^{exp}$, together with their uncertainties
$\delta C^{exp}$ (Tab. I in \cite{Radecky:1981fn}). These correction factors are related to detector efficiencies and multiple kinematic cuts. 
We define estimator for both neutron channels as:
\begin{widetext}
\begin{equation}\label{eq:chi2anl23}
\chi^2_{A2,3} = \hspace{-4pt}\sum_{i=1}^{N_{A2,3}} \hspace{-2pt}\frac{\displaystyle \left(
\frac{\sigma_{A2,3;i}^{th}}{\sigma_{A2,3,tot}^{th}}
N_{A2,3}^{exp}  p_{ANL}
-\hspace{-1pt} N_{A2,3;i}^{exp}\hspace{-1pt}\right)^2}
{N_i^{exp}}\hspace{-3pt} +  \left(\frac{\displaystyle \frac{\sigma_{A2,3,tot}^{th}}
{\sigma_{A2,3,tot}^{exp}\cdot p_{ANL}} -1}{\Delta p_{ANL}}\right)^2,
\end{equation}
\end{widetext}
where 
\begin{eqnarray}
N_{A2,3}^{exp}     &=& \sum_{j=1}^{12}  N_{A2,3;j}^{exp} \\
\sigma_{A2,3,tot}^{th} &=& \sum_{j=1}^{12} \sigma_{A2,3;j}^{th}\\
\end{eqnarray}
$\sigma_{A2,3,tot}^{exp}$ is the total cross section for $A2,3$ channel, $N_{A2,3;i}^{exp}$ is the number of events in the $i$-th bin of the $A2,3$ channels.

Some of the experimental bins contain too few events for a $\chi^2$-based analysis. We have combined some of the neighbouring bins 
in order to keep a meaningful event statistics and the number of $Q^2$ bins is $12$ in both neutron channels.
The upper bound on neutrino energy is now $E_{max}=1.5\;$GeV and one has to account 
for that fact by changing the integration limits and normalization
factor in Eq. (\ref{eq:sigmath}).

Eventually, the complete $\chi^2$-function for the ANL data reads,
\begin{equation}
\chi^2_{ANL} = \sum_{k=1}^3\chi^2_{Ak}.
\end{equation}

\section{Results}
\label{sec:results}

\subsection{Electromagnetic fits}

The best fit results of our vector form factor parametrization given by Eqs. (\ref{eq:Kellyc5}-\ref{eq:c5K}) are shown in Table \ref{tab:ourfitt}. 
For our best fit the value of $C_3^V(0)$ is close to the one from Ref. \cite{Lalakulich:2006sw} and we get a clear beyond-dipole $Q^2$ dependence of
$C_3^V(Q^2)$ and $C_4^V(Q^2)$. Surprisingly, the $Q^2$ dependence of $C_5^V(Q^2)$ is exactly dipole $(1+Q^2/M_V^2)^{-2}$
with $M_V=0.84\;$GeV being the standard vector mass.

\begin{table}[htpb]
\centering
\caption{Best fit coefficients for vector form factors given by Eqs. (\ref{eq:Kellyc5}-\ref{eq:c5K}) to be used in neutrino scattering data analysis.
We do not report $1 \sigma$ errors because of hybrid character of our estimator, see explanations in the text.}
\begin{tabular}{|cccccccc|}
\hline
\hline
 &  &  &  &  &   &  &  \\
 $C_3^V(0)$ & $C_5^V(0)$ & $A$ & $B$ & $C$ & $D$ & $K_1$ & $K_2$\\
 &  &  &  &  &  &  &   \\
\hline
 &  &  &  &  &  &  &   \\
$\, 2.10$
& $\, 0.63$
& $\, 4.73$
& $-0.39$
& $\,5.59$
& $\, 1.00$
& $\, 0.13$
& $\, 1.68$\\
 &  &  &  &  &  &  &  \\
\hline
\hline
\end{tabular}
\label{tab:ourfitt}
\end{table}

\begin{figure}[htb]
\centering\includegraphics[width=\columnwidth]{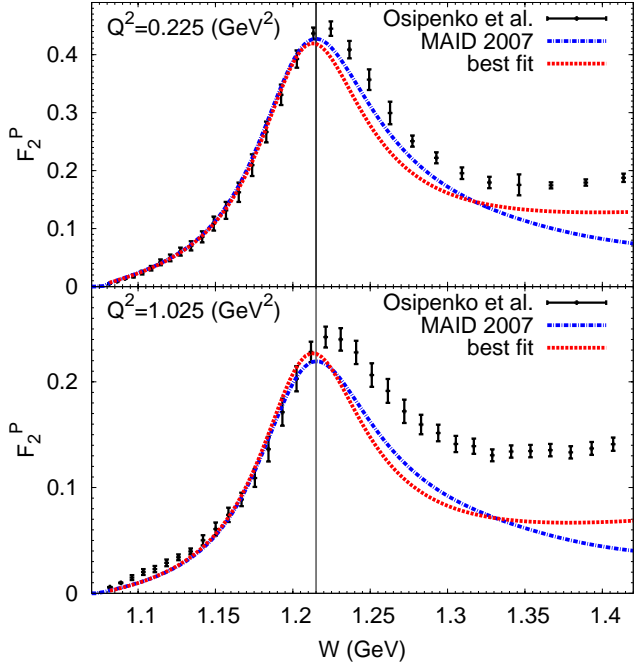}
\caption{(Color online) Best fit results for vector form factors given by Eqs. (\ref{eq:Kellyc5}-\ref{eq:c5K}) plotted against experimental data
from Ref. \cite{Osipenko:2003ua} as well as MAID2007 predictions. Vertical lines show the $2\pi$ production threshold.}\label{fig:ourfit1}
\end{figure}

Fig. \ref{fig:ourfit1} shows that qualitatively in the
region below two pion production threshold our fit reproduces the data rather well.  
In the same figure we show also predictions from the MAID2007 model.
In order to compare both results we calculated the $\chi^2$ contribution from data points below the $2\pi$ threshold ($\chi^2_{<2\pi}$).
The same $\chi^2$ function with the MAID2007 model predictions gives $\chi^2_{W<M_p+2m_\pi}/NDF\approx 12.1$
and with our best fit results-$\chi^2_{W<M_p+2m_\pi}/NDF\approx 13.6$.
Our form factors lead to better agreement with the electron scattering data than the form factors considered in Ref. \cite{Lalakulich:2006sw} (with the same HNV background model)
giving $\chi^2_{W<M_p+2m_\pi}/NDF =16.5$.
Inspection of Fig. \ref{fig:ourfit1} (and also similar figures not shown in the paper) shows that most of the contribution
to $\chi^2$ comes from a region of low $W$.
Our fits are going to be used in the analysis of
neutrino scattering data and some discrepancy at low $W$ is of no practical importance.

\begin{figure*}[htb]
\centering\includegraphics[width=0.7\textwidth]{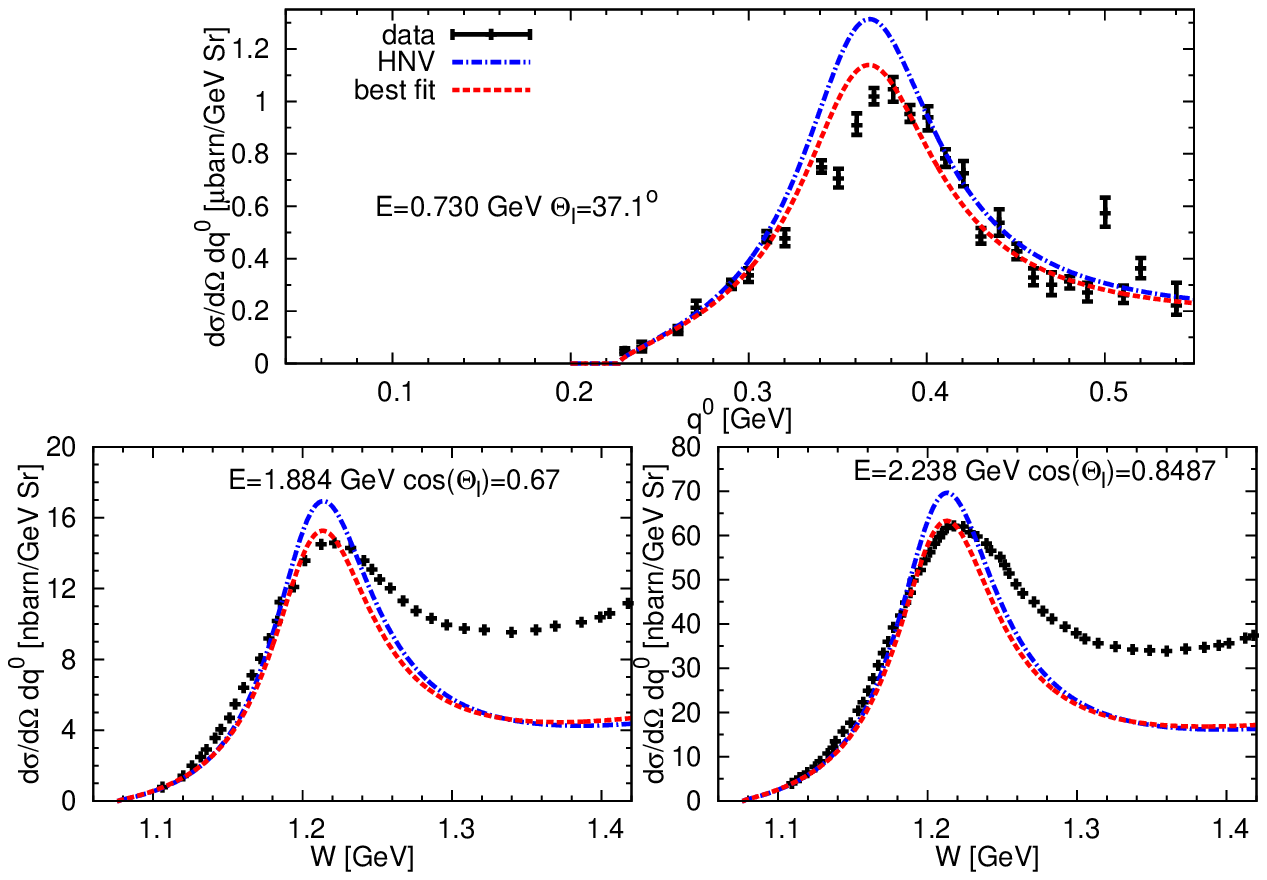}\caption{(Color online) Comparison of our best fit and HNV model with
Lalakulich-Paschos form factors of Ref. \cite{Lalakulich:2006sw} plotted against inclusive $p(e,e')$ data (not included in the fit) from Ref.
\cite{O'Connell:1984nw} (top) and Ref. \cite{Christy2007} (bottom). The $Q^2$ values at peak are from top to bottom and left to right:
$0.1\, \mathrm{(GeV^2)}$, $1.15\, \mathrm{(GeV^2)}$ and $0.95\, \mathrm{(GeV^2)}$ respectively.} \label{fig:elprod}
\end{figure*}

Fig. \ref{fig:elprod} shows an example of the performance of our best fit and form factors from Ref. \cite{Lalakulich:2006sw} with the same background. Our fmodel gives results closer to the experimental data than the form factors proposed in Ref. \cite{Lalakulich:2006sw}.

\subsection{Axial fits}

\begin{figure}[htb]
\centering\includegraphics[width=\columnwidth]{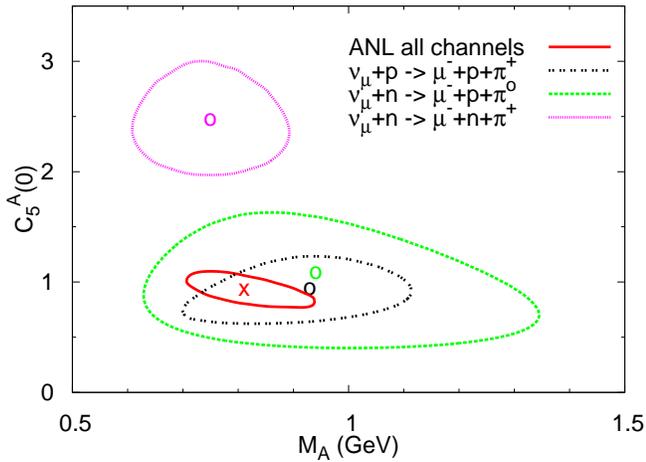}
\caption{(Color online) 1$\sigma$ uncertainty contours for fits on free target. } \label{fig:sigfree}
\end{figure}

\begin{figure}[htb]
\centering\includegraphics[width=\columnwidth]{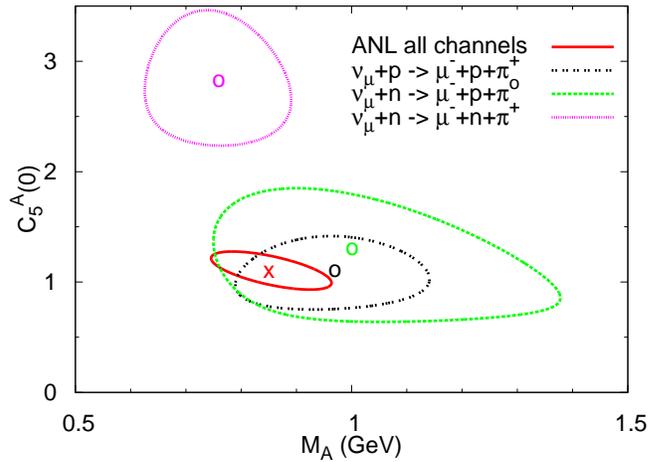}\caption{(Color online) 1$\sigma$
uncertainty contours for fits on deuteron target.} \label{fig:sigdeut}
\end{figure}

For the axial contribution to $N\to\Delta$ transition our analysis assumes, that $C_5^A(0)$, $M_{A\Delta}$ and 
normalization factor $p_{ANL}$ are free fit parameters. We present our results in Tab. \ref{tab:axfit} and in Figs. \ref{fig:sigfree} and \ref{fig:sigdeut}.

\begin{table}[htpb]
\caption{Best fit for the $\Delta$(1232) axial form factors. Upper table: free nucleon target, lower table: deuteron target.
 1$\sigma$ contours for physical parameters can be found in Figs. \ref{fig:sigfree} and \ref{fig:sigdeut}.
 Errors for $C_5^A(0)$ and $M_{A\Delta}$ where obtained after marginalization of $p_{ANL}$. }
\label{tab:axfit}\centering
\begin{tabular}{|c|c|ccccc|}
\hline
\hline
& &  &  &  & &  \\
&Fit & $C_5^A(0)$ & $M_{A\Delta}$(GeV) & $p_{ANL}$ & $\chi^2/NDF$ & $NDF$ \\
& &  &  &  & &  \\
\hline
& &  &  &  & &  \\
\multirow{4}{*}{\rotatebox[origin=c]{90}{Free n+p}}&A1 & 0.94$\foot{+0.30}{-0.30}$ & 0.93$\foot{+0.18}{-0.19}$ & 1.03 & 0.15 &6  \\
&A2 & 1.09$\foot{+0.50}{-0.69}$ & 0.94$\foot{+0.30}{-0.30}$ & 0.93 & 1.55 &9 \\
&A3 & 2.48$\foot{+0.52}{-0.52}$ & 0.75$\foot{+0.14}{-0.14}$ & 0.94 & 1.56 & 9 \\
& &  &  &  & &  \\
&Joint & 0.93$\foot{+0.13}{-0.13}$ & 0.81$\foot{+0.09}{-0.09}$ & 0.89 & 2.11 & 30\\
& &  &  &  & &  \\
\hline
& &  &  &  & &  \\
\multirow{4}{*}{\rotatebox[origin=c]{90}{Deuteron}}& A1 & 1.11$\foot{+0.32}{-0.34}$ & 0.97$\foot{+0.17}{-0.17}$ & 1.04 & 0.20 &6 \\
&A2 & 1.31$\foot{+0.49}{-0.77}$ & 1.00$\foot{+0.27}{-0.25}$ & 0.93 & 1.52 & 9 \\
&A3 & 2.83$\foot{+0.62}{-0.60}$ & 0.76$\foot{+0.13}{-0.13}$ & 0.94 & 1.47 & 9 \\
& &  &  &  & &  \\
&Joint & 1.10$\foot{+0.15}{-0.14}$ & 0.85$\foot{+0.09}{-0.08}$ & 0.90 & 2.06& 30 \\
& &  &  &  & &  \\
\hline
\hline
\end{tabular}
\end{table}

In the Tab. \ref{tab:axfit} are the results for fits to all three channels separately, and also the joint fit to three channels together.
In each case the number of degree of freedom is calculated as:\\ $NDF=$ No. $Q^2$ bins $-$ No. fitted parameters.\\
In order to illustrate a role of deuteron effects we show also the results for a ``model'' of deuteron as consisting from free proton and neutron. 

In both free target and deuteron target cases we see, that taken separately the $p\pi^+$ (A1) and $p\pi^0$ (A2) channels are statistically
consistent, albeit their
predicted scale parameters differ by around 10\%. The latter channel seems to carry less information on the $N\to \Delta$ 
transition axial current
than the first one, which
is reflected in larger uncertainty contours. This could be explained by a bigger background contribution to that channel,
which makes it less sensitive to changes in the
$\Delta$ resonance description.

\begin{figure}[htb]
\centering\includegraphics[width=\columnwidth]{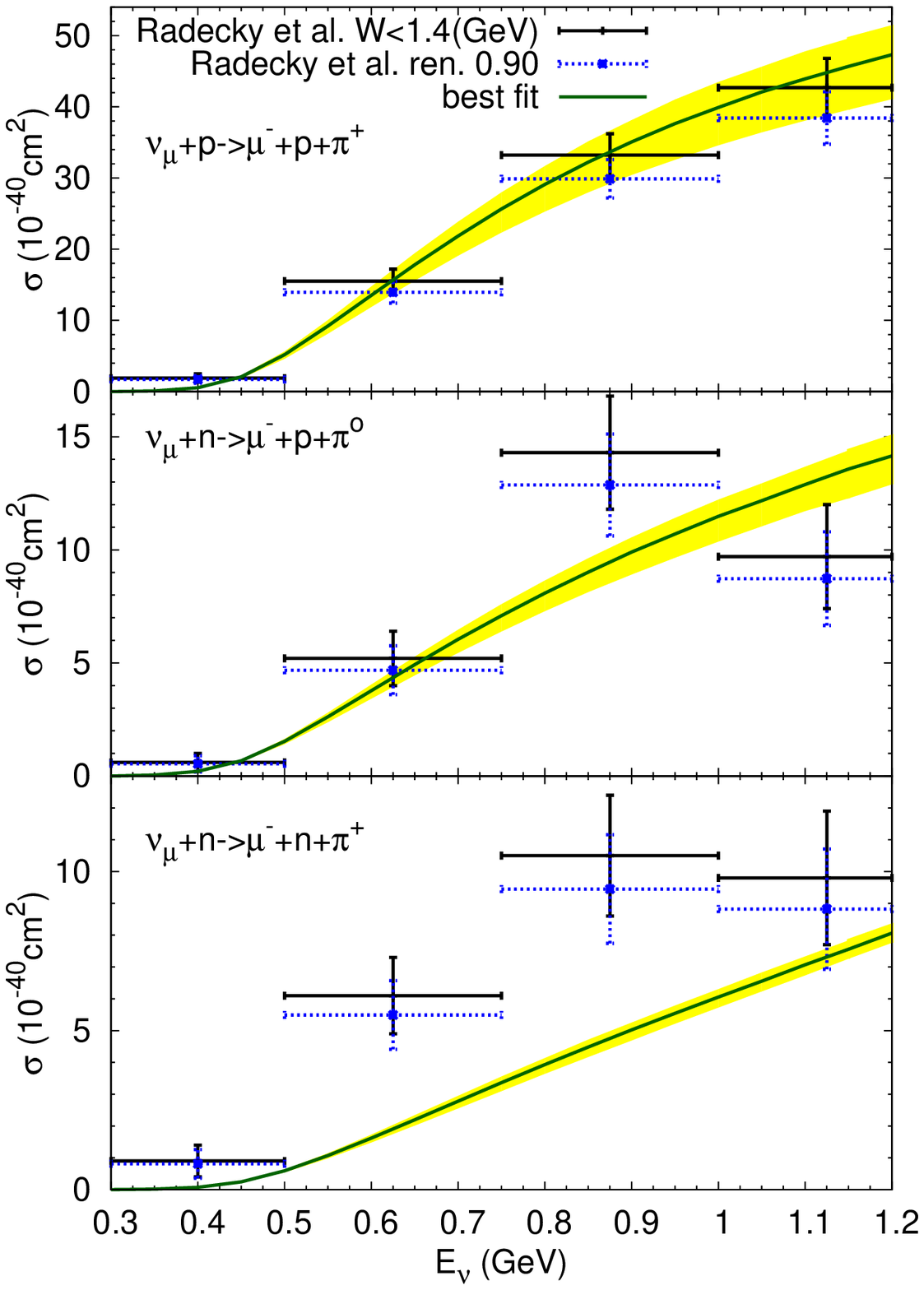}\caption{(Color online) Total cross sections 
for our best fit form factors on deuteron target with 1$\sigma$ error bands.
Black/solid lines represent the experimental data from Ref. \cite{Radecky:1981fn}, blue/dotted lines - the same data
multiplied by the best fit value of $p_{ANL}=0.90$ (see Tab. \ref{tab:axfit}).} \label{fig:totals}
\end{figure}

The biggest difficulty is encountered in the $n\pi^+$ (A3) channel, where we obtain $C_5^A(0)$ twice as large
as for the other two channels and $M_{A\Delta}$ significantly smaller. Here the number of events reported by ANL is comparable
to $p\pi^0$ channel, but theoretical cross section predictions with nonresonant background
are smaller, as one can readily see in the Fig. \ref{fig:totals}. This results in the drastic overestimation of $C_5^A(0)$.
Still, the fits to separate isospin channels give acceptable values of $\chi^2_{min}$ for both  neutron channels.

Deuteron effects affect mostly the value of $C_5^A(0)$, by up to 20\% depending on interaction channel.
The same applies to the joint fit. A significant improvement with respect to previous fits to HNV model done in
Refs. \cite{Hernandez:2007qq,Hernandez:2010bx}
is that with deuteron target effects we get the best fit value of $C_5^A(0)$ within 1$\sigma$ range from the
theoretical Goldberger-Treiman relation.
The joint fit agrees also on the 1$\sigma$ level with separate fits on $p\pi^0$ and $p\pi^+$ channels.
Deuteron effects lead to a slight improvement in the values of $\chi^2_{min}$.

\begin{figure}[htb]
\centering\includegraphics[width=\columnwidth]{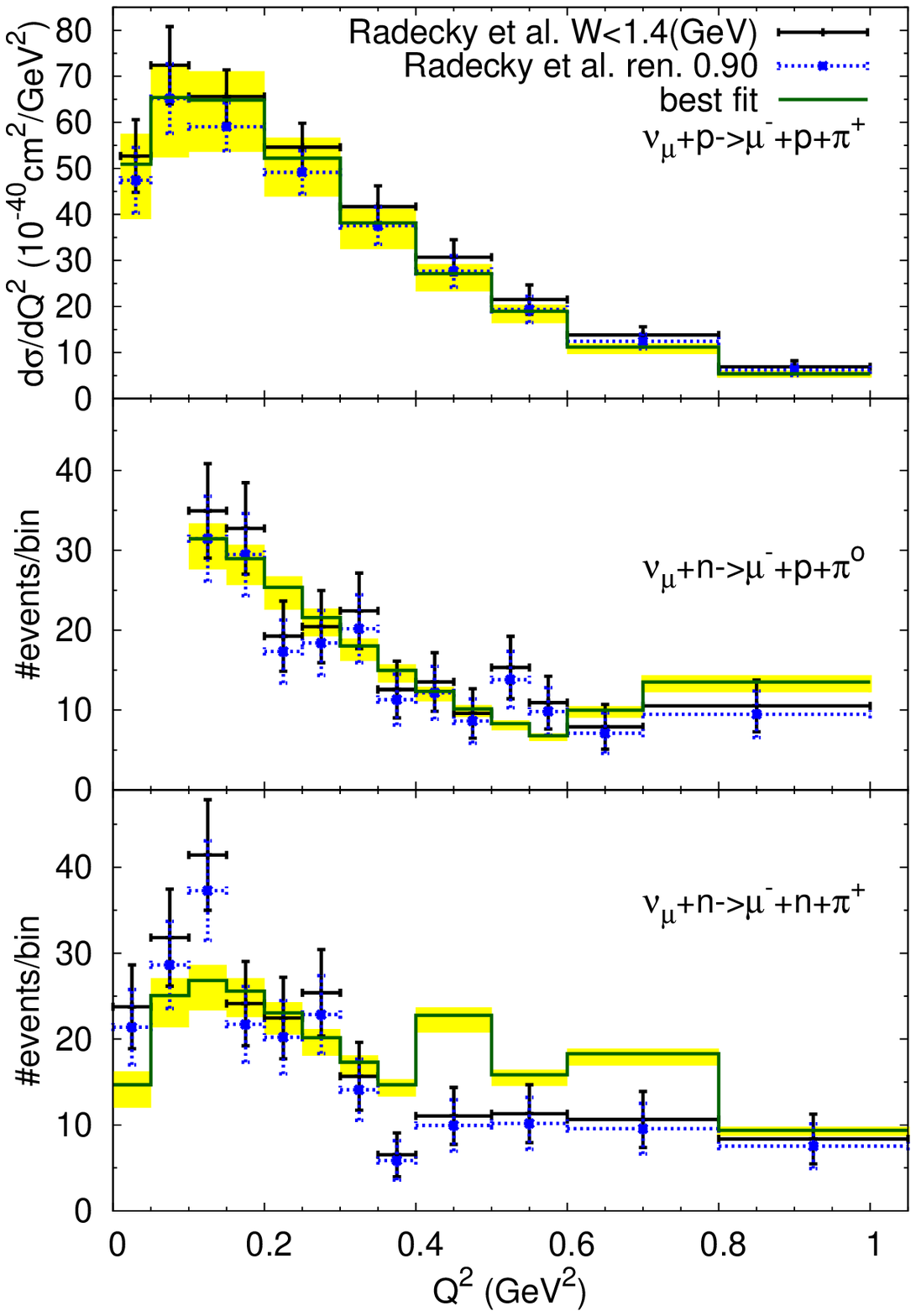}
\caption{(Color online) The ANL $Q^2$ cross section and event distributions for our best fit and deuteron target.
Black/solid lines represent the experimental data from Ref. \cite{Radecky:1981fn}, blue/dotted lines- the same data
multiplied by the best fit value of $p_{ANL}=0.90$ (see Tab. \ref{tab:axfit}).} \label{fig:dq2}
\end{figure}

We have compared total cross section and $Q^2$ event distribution from the ANL experiment and our best fit.
They are presented in Fig. \ref{fig:totals} and in Fig. \ref{fig:dq2} respectively.
They reflect previously described problems with the $n\pi^+$ channel. For two other channels
we get a good agreement with the data.

\begin{figure}[htb]
\centering\includegraphics[width=\columnwidth]{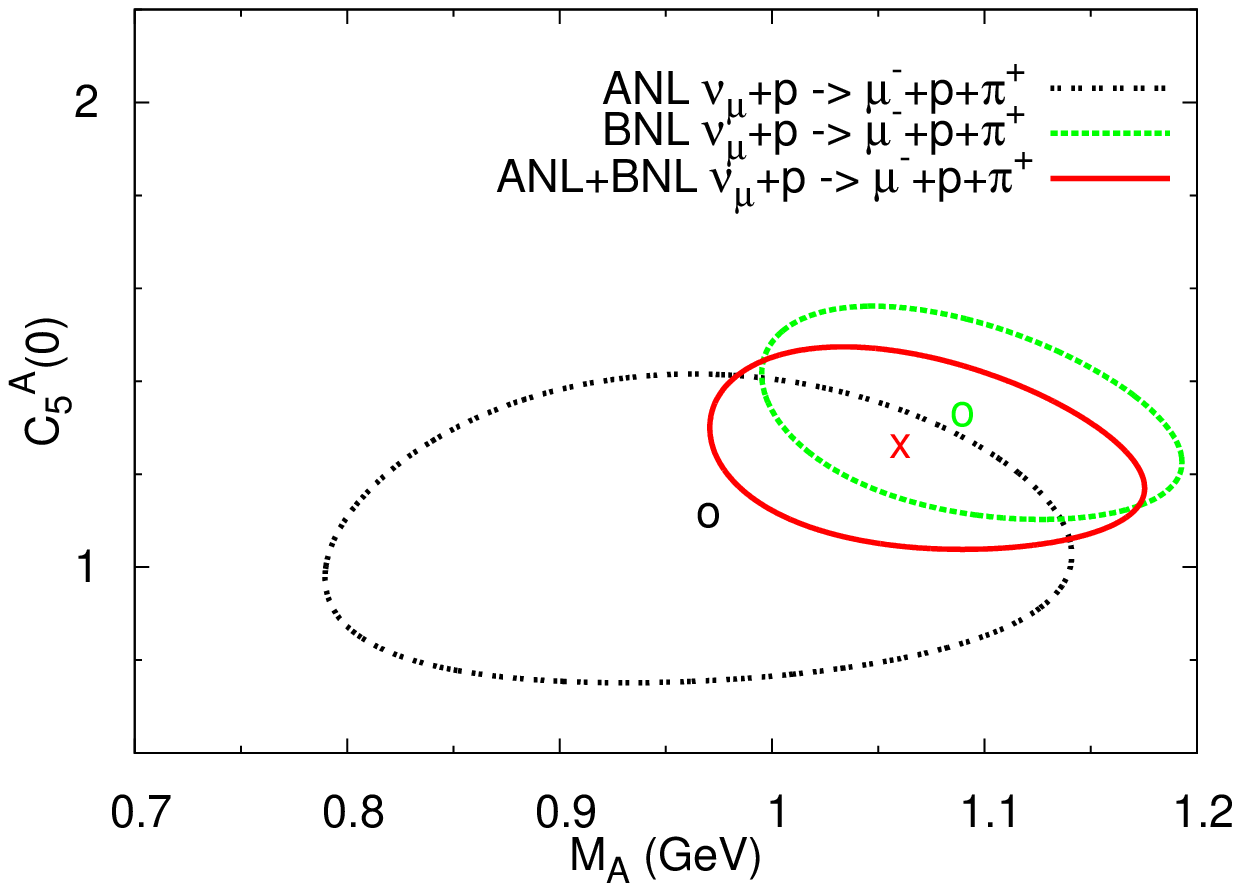}\caption{(Color online) Best fit results for the ANL+BNL $p\pi^+$ data 
on deuteron target
with 1$\sigma$ error bands.} \label{fig:bnlfit}
\end{figure}

Fitted normalizations factors $p_{ANL}$ are different for neutron and proton channel 
as long as one considers separate fits. The proton channel prefers the data to be scaled up and both neutron channels 
prefer the data to be scaled down. 
Inclusion of deuteron effects does not change the value of fitted $p_{ANL}$. The joint fit uses the same $p_{ANL}$ parameter
for all channels and seems to prefer the data to be scaled down even more ($p_{ANL}\approx 0.90$ both for free and deuteron targets).
These values of $p_{ANL}$ are all well within the assumed error $\Delta p_{ANL}$. This indicates 
that our fitting procedure is numerically stable. The effect of the fitted overall normalization factor
has been shown in Fig. \ref{fig:totals} for the total cross sections and in Fig. \ref{fig:dq2} for the differential cross sections.

Finally, we noticed that the best fit values for $C_5^A(0)$ and $M_{A\Delta}$ 
are different from those obtained in Ref. \cite{Graczyk:2009qm} because in the 
current analysis the non-resonant background contribution is included.
\subsection{Inclusion of BNL data}

We repeated the similar analysis with the BNL SPP data
published in \cite{Kitagaki:1986ct} and  \cite{Kitagaki:1990vs}. The BNL neutrino flux was of somewhat higher energy than ANL,
$\left\langle E\right \rangle \approx 1.6\;$GeV,
with flux uncertainty $\Delta p_{BNL}\approx 10\%$, see Ref. \cite{Graczyk:2009qm}. For our purposes, 
the most useful data is for the $\nu_\mu + p \to \mu^- + p + \pi^+$ reaction in a form of distribution of events with a cut $W<1.4$~GeV. 
Neutron channel results have been reported without a $W$ cut and they contain a large contamination coming from heavier resonances. We used 
the same formula for $\chi^2$  function as in Ref. \cite{Graczyk:2009qm}, Sec. 5.2
with the normalization factor for the BNL data, $p_{BNL}$, treated as a free fit parameter.

The joint ANL+BNL data fit was done for the $p\pi^+$ channel and the best fit result is: $C_5^A(0)=1.26\foot{+0.20}{-0.21}$
(consistent with Goldberger-Treiman relation) and $M_{A\Delta}=1.06\foot{+0.10}{-0.09}\;$GeV ($\chi^2/NDF=0.74$ with $NDF=35$). Our results are different from those obtained
in Ref. \cite{Hernandez:2010bx} because both studies use distinct estimators $\chi^2$. In Ref. \cite{Hernandez:2010bx}
only total cross sections information from the BNL data is utilized. As explained above, we have used an information from the shape of 
$Q^2$ event distributions as well. In the study in Ref. \cite{Hernandez:2010bx} most of the data points come
from the ANL experiment and joint best fit value of $C_5^A(0)$ becomes smaller.

We have obtained slightly different results from the ones from 
Ref. \cite{Graczyk:2009qm} ($C_5^A(0)=1.19\foot{+0.08}{-0.08}$ and $M_{A\Delta}=0.94\foot{+0.03}{-0.03}\;$GeV), where the same
$\chi^2$ definition was used. 
The reasons are that in the current study:
\begin{itemize}
\item We included the nonresonant background.
\item We used new vector form factors.
\item We used a better description of deuteron effects. In Ref. \cite{Graczyk:2009qm} an effective
treatment of deuteron effects based on \cite{AlvarezRuso:1998hi} was applied.
\end{itemize}

\section{Conclusions}
\label{sec:conclusions}

In this paper we made a new attempt to get an information about weak $N\to\Delta$ transition
matrix elements. We first introduced new vector form factors, consistent 
with the HNV model of the nonresonant background.
In the next step we investigated all three neutrino-free nucleon SPP channels, most importantly also neutrino-neutron channels that were never before
used in the phenomenological studies.

Our main result is that the obtained value of $C_5^A(0)$ agrees, on the 1$\sigma$ level,
with the Goldberger-Treiman relation.
Also, our results confirm that there is a strong tension between $n\pi^+$ and remaining two channels in the sense that the same theoretical
model does not seem to reproduce all the data in a consistent way.

There can be various reasons for that, some of them have been already mentioned:

\begin{itemize}
\item ANL data for the neutron SPP channels is of poor statistics.
 \item The HNV model for the background is well justified only near the pion production threshold and perhaps it is not
 reliable in the $\Delta(1232)$ peak region.
\end{itemize}

Still another reason of theoretical difficulties may come from a missing unitarization of the model. The unitarity constraint, following
the Watson theorem \cite{Watson:1952ji}, imposes a relation between phases in weak neutrino-nucleon and pion-nucleon elastic scattering amplitudes
not satisfied in our approach. 
In a recent study Nieves, Alvarez-Ruso, Hernandez and Vicente-Vacas \cite{nieves_nuint14} tried to correct the HNV model by introducing 
phenomenological phases in the leading multipole amplitude. This approach leads to a better agreement of the obtained 
best fit value of $C_5^A(0)$ with the Goldberger-Treiman relation. More theoretical studies in this direction are necessary.

Another observation is that better statistics SPP measurements in the $\Delta$ region on proton or deuteron targets are badly needed. Keeping
in mind difficulties in the treatment of nuclear effects on heavier targets it is the only way to get precise information about the
$N\to \Delta$ axial transition matrix elements. 

\section*{Acknowledgements}
We thank Luis Alvarez-Ruso for fruitful discussions.

JTS and JZ were supported by Grant 4585/PB/IFT/12 (UMO-2011/M/ST2/02578).

Numerical calculations were carried out in Wroclaw Centre for
Networking and Supercomputing 
(\url{http://www.wcss.wroc.pl}),
grant No. 268.

\bibliographystyle{apsrev4-1}

\bibliography{bibdrat}
\end{document}